\documentclass[twocolumn]{aastex63}
\usepackage{amsmath}

\def\ltsima{$\; \buildrel < \over \sim \;$}
\def\simlt{\lower.5ex\hbox{\ltsima}}
\def\gtsima{$\; \buildrel > \over \sim \;$}
\def\simgt{\lower.5ex\hbox{\gtsima}}
\def\g{{\rm\,g}}

\def \r {r$^{1/4}$ }


\def\s{\ifmmode \widetilde \else \~\fi}
\def\={\overline}

\def\spose#1{\hbox to 0pt{#1\hss}}

\def\lta{\mathrel{\spose{\lower 3pt\hbox{$\mathchar"218$}}
     \raise 2.0pt\hbox{$\mathchar"13C$}}}
\def\gta{\mathrel{\spose{\lower 3pt\hbox{$\mathchar"218$}}
     \raise 2.0pt\hbox{$\mathchar"13E$}}}
\def\Dt{\spose{\raise 1.5ex\hbox{\hskip3pt$\mathchar"201$}}}    
\def\dt{\spose{\raise 1.0ex\hbox{\hskip2pt$\mathchar"201$}}}    

\def\r{${\rm r^{1/4}}$}

\def\dotsfill{\leaders\hbox to 1em{\hss.\hss}\hfill}
\def\sun{\odot}
\def\earth{\oplus}

\def\ltsima{$\; \buildrel < \over \sim \;$}
\def\gtsima{$\; \buildrel > \over \sim \;$}
\def\lsim{\lower.5ex\hbox{\ltsima}}
\def\gsim{\lower.5ex\hbox{\gtsima}}
\def\lapp{\ifmmode\stackrel{<}{_{\sim}}\else$\stackrel{<}{_{\sim}}$\fi}
\def\gapp{\ifmmode\stackrel{>}{_{\sim}}\else$\stackrel{<}{_{\sim}}$\fi}

\defcitealias{V21}{V21}
\usepackage{subfigure}

\submitjournal{ApJ}

\shorttitle{Revisiting a disky origin for the faint branch of the Sgr stream}
\shortauthors{Oria et al.}
\graphicspath{{./}{figures/}}

\begin{document}

\title{Revisiting a disky origin for the faint branch of the Sagittarius stellar stream}

\correspondingauthor{Pierre-Antoine Oria}
\email{pierre-antoine.oria@astro.unistra.fr}

\author{Pierre-Antoine Oria}
\affiliation{Universit\'e de Strasbourg, CNRS, Observatoire astronomique de Strasbourg, UMR 7550, F-67000 Strasbourg, France}

\author{Rodrigo Ibata}
\affiliation{Universit\'e de Strasbourg, CNRS, Observatoire astronomique de Strasbourg, UMR 7550, F-67000 Strasbourg, France}

\author{Pau Ramos}
\affiliation{Universit\'e de Strasbourg, CNRS, Observatoire astronomique de Strasbourg, UMR 7550, F-67000 Strasbourg, France}

\author[0000-0003-3180-9825]{Benoit Famaey}
\affiliation{Universit\'e de Strasbourg, CNRS, Observatoire astronomique de Strasbourg, UMR 7550, F-67000 Strasbourg, France}

\author{Raphaël Errani}
\affiliation{Universit\'e de Strasbourg, CNRS, Observatoire astronomique de Strasbourg, UMR 7550, F-67000 Strasbourg, France}

\begin{abstract}
We investigate ways to produce the bifurcation observed in the stellar stream of the Sagittarius dwarf galaxy (Sgr). Our method consists in running $N$-body simulations of Sgr falling into the Milky Way for the last 3~Gyr, with added test particles on disk orbits that span a wide range of initial positions, energies, and angular momenta. We find that particles that end up in the faint branch are predominantly high angular momentum particles that can all originate from a single plane within the progenitor, nearly perpendicular both to the orbital plane of the progenitor and to the Milky Way stellar disk. Their original configuration at the start of the simulation corresponds to spiral features already present 3~Gyr ago, which could be, {\it e.g.}, the result of a disk-like component being tidally perturbed, or the tidal tails of a satellite being disrupted within Sgr. We then run a simulation including the self-gravity of this disky component. Despite the remaining ambiguity of its origin, this disk component of the Sgr dwarf with spiral over-densities provides a first step towards a working model to reproduce the observed faint branch of the bifurcated Sgr stream. 
\end{abstract}


\section{Introduction}
\label{sec:Introduction}

Since its discovery \citep{1994Natur.370..194I,1995MNRAS.277..781I}, the Sagittarius dwarf galaxy (Sgr) has been under intense scrutiny as the closest example of an on-going galactic merging event. The stellar stream generated by its tidal disruption \citep{2001ApJ...547L.133I,2003ApJ...599.1082M} is an extended and complex kinematic structure in the stellar halo of the Milky Way (MW), and as such constitutes an invaluable source of information on the gravitational potential and history of {\it both} the MW and the progenitor dwarf galaxy itself.

Over the years, several models have been put forward in order to reproduce the shape of the stream and its kinematics (e.g. \citealt{1997AJ....113..634I,2004ApJ...610L..97H,2005ApJ...619..807L}). Among those, \citet{2010ApJ...714..229L} reproduced most of the observational constraints at the time, involving however an unrealistic and unstable triaxial dark matter halo configuration for the MW. The latest up-to-date model by-passing this problem is that of \citet[][hereafter \citetalias{V21}]{V21}, in which the Sgr dwarf is infalling in the joint evolving gravitational potential of the MW and the Large Magellanic Cloud (LMC), yielding a very good agreement with recent Gaia data.

One of the remaining mysteries about the Sgr stream is the presence of a bifurcation, in the form of a faint branch running parallel to the main brighter branch, observed first in the leading arm \citep{2006ApJ...642L.137B}, then in the trailing arm \citep{2012ApJ...750...80K}. More recently, this bifurcation has been outlined with great precision by \citet{2021arXiv211202105R} using the latest Gaia EDR3 data \citep{2021A&A...649A...1G}. 

\citet{2006ApJ...651..167F} proposed an early explanation for the bifurcation, as the result of the young leading, old leading and trailing wraps overlapping and being slightly displaced due to the precession of the orbit \citep{2009ApJ...700.1282Y,2010ApJ...712..516N}, but this model did not match later observations of the stream.

\citet{2010MNRAS.408L..26P} then proposed a model in which the Sgr dwarf originally consisted of a rotating stellar disk embedded in a cold dark matter halo. A disk slightly misaligned with respect to the orbital plane was shown to produce a bifurcation as observed in the Sgr stream. However, the model predicted some remnant rotation in the centre of Sgr today, which was not observed \citep{2011ApJ...727L...2P}. 

Although not in the context of the bifurcation, an originally disky Sgr was also studied by \citet{2010ApJ...725.1516L} in order to explain the elongated shape of the remnant. This model makes use of the tidal stirring mechanism \citep{2001ApJ...547L.123M,2011ApJ...726...98K} according to which dwarf spheroidal galaxies are the outcome of disky satellite galaxies being deformed due to galactic tides.

In this letter, we re-investigate the production of a bifurcation by selecting, within simulations of the Sgr stream, particles that end up in the observed faint branch, and then examine the properties of the initial conditions. 

\section{Methodology}
\label{sec:Methodology}

\subsection{Reference model} The underlying model that we use for the present work is the $N$-body simulation proposed by \citetalias{V21} of the Sgr dwarf falling into a joint, evolving MW and LMC gravitational potential. This model constitutes an ideal basis for our investigations as it already reproduces many observational constraints (e.g. positions, proper motions, distances and line-of-sight velocities, with the inclusion of the LMC being key for the latter two, especially in the leading arm), leaving us free to focus our efforts on the production of the bifurcation. In this context, the simulation starts 3~Gyr ago, at which point Sgr is made of a spherical King distribution stellar component of mass $2\times 10^8$ M$_{\odot}$, immersed in a spherical dark halo of mass $3.6\times 10^9$ M$_{\odot}$. The stellar and dark matter components are made of $2\times 10^5$ particles each. The MW and LMC models are described in detail in \citetalias{V21}.

\subsection{Sagittarius model and simulation} Using the $N$-body code \textsc{Gyrfalcon} \citep{2000ApJ...536L..39D}, we first reproduce the simulation provided by \citetalias{V21}. Then, we add test particles to the initial conditions of the Sgr dwarf (self-gravity will be included in \S\ref{sec:self_gravity}) to see which ones are more likely to end up in the faint branch of the Sgr stream by the end of the simulation.

Given the aforementioned works hinting strongly at the importance of rotation in Sgr to produce a bifurcation, we choose to populate our simulations with test particles with wide ranges of angular momenta. The sample of test particles is produced by generating stellar disks using \textsc{Agama} \citep{2019MNRAS.482.1525V} and giving them each a different inclination w.r.t. the orbital plane of Sgr. Each disk has a scale radius $R_{\rm disk}=0.9$ kpc (\citetalias{V21}'s King model has a scale radius of $1$ kpc), a scale height $H_{\rm disk}=0.18$ kpc, central velocity dispersion $\sigma_{r,0}=4$ kms$^{-1}$, and is generated through a QuasiIsothermal distribution function. The full \textsc{Agama} script for generating the disk is available in the shared data.

We use a right-handed Cartesian coordinate system centered on the MW with the $xy$ plane being its disk plane, and the $x$-axis pointing along the Sun-Galactic center direction, with the sun at $(x,y,z)=(-8,0,0)$ kpc. Our disks are generated in this plane, then we rotate them before launch by probing inclination angles $i$ every $20^{\circ}$, both around the $x$-axis and around the $y$-axis using the following matrices respectively:
\begin{align}
\label{eq:rotation_matrices}
  & R_{x} = 
  \begin{pmatrix}
    1 & 0 & 0 \\
    0 & \cos(i) & -\sin(i) \\
    0 & \sin(i) & \cos(i)
  \end{pmatrix},\\
  & R_{y} = 
   \begin{pmatrix}
    \cos(i) & 0 & -\sin(i) \\
    0 & 1 & 0 \\
    \sin(i) & 0 & \cos(i)
  \end{pmatrix}.
\end{align}
For reference, for such a disk to be in the orbital plane of Sgr (at present), it would have to be rotated around the $x$-axis with the $R_x$ matrix by an angle of  $i\simeq100^{\circ}$. Before applying a rotation, our disk models have angular momentum aligned with the positive $z$-axis. Preliminary tests showed us that particles in the inner regions of the disk would not end up in the faint branch, but rather end up close to the remnant of the progenitor. This is understandable as those particles are deeper in the potential well of the King model and much harder to strip. We thus select from those disks the $2\times10^4$ outermost test particles out of the $5\times10^4$ particles generated, allowing us to better probe the regions of interest. This corresponds to a hole in the inner $\simeq1.5$ kpc of the disks. 

\subsection{Stream selection} 

\begin{figure*}[!htb]
\centering
  \includegraphics[angle=0,  clip, width=\textwidth]{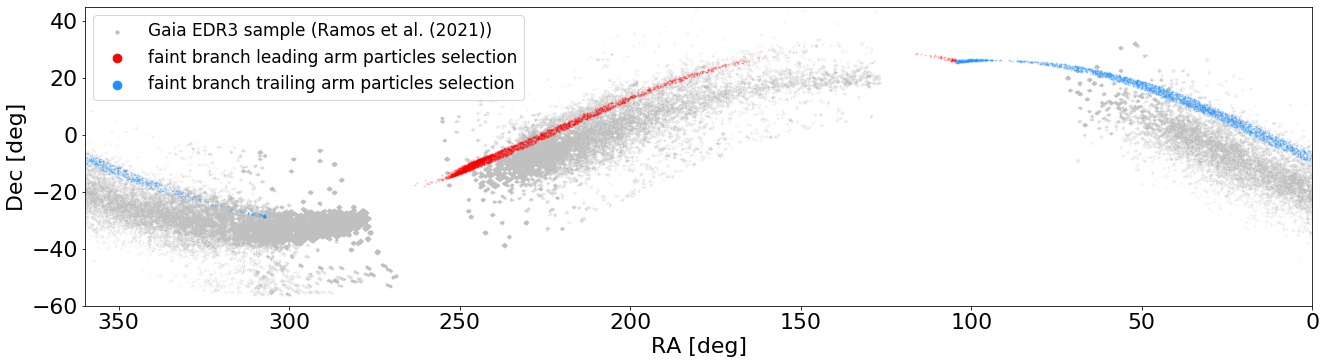}
   \caption{The Gaia EDR3 Sgr sample from \citet{2021arXiv211202105R} in the background (grey), with our selection of faint branch test particles from our simulations from Section~\ref{subsec:yaxis} over-plotted (red for leading arm, blue for trailing arm). The remnant of the progenitor lies in the $275\leq\rm{RA}\leq300$ region.}
\label{fig:gaia_plus_test_particles}
\end{figure*}

In order to pick the particles that best match the observations of the faint branch of the Sgr stream, we use the polynomial fits proposed by \citet[Table 1]{2021arXiv211202105R} for the ($\tilde\Lambda_\odot$,$\tilde \beta_\odot$) coordinate system centred on Sgr, introduced in \citet{2003ApJ...599.1082M} and representing the latitude and longitude along its stream. We use the slight sign modification of \citet{2014MNRAS.437..116B} for this coordinate system, in which $\tilde\Lambda_\odot$ increases towards the leading arm. In the final snapshot of our simulations, we thus select the test particles with $\tilde \beta_\odot$ such that $|\tilde \beta_\odot/P(\tilde\Lambda_\odot)-1|<0.2$ where
\begin{equation}
\label{eq:polynomial_leading}
P(\tilde\Lambda_\odot)=-0.0003819\tilde\Lambda_\odot^2+0.01904\tilde\Lambda_\odot+6.084
\end{equation}
applies to the leading arm part of the faint branch, and %
\begin{equation}
\label{eq:polynomial_trailing}
P(\tilde\Lambda_\odot)=-0.001563\tilde\Lambda_\odot^2-0.2805\tilde\Lambda_\odot-3.040
\end{equation} 
applies to the trailing arm part of the faint branch. We also require that $|\tilde\Lambda_\odot|\geq20$ in order to exclude the progenitor. In Figure~\ref{fig:gaia_plus_test_particles}, we show the Gaia EDR3 Sgr sample from \citet{2021arXiv211202105R}, and we over-plot what our faint branch selection of Section~\ref{sec:test_particles} based on Equations~\eqref{eq:polynomial_leading} and~\eqref{eq:polynomial_trailing} from our simulations with test particles looks like.

\section{Results}
\label{sec:Results}
\subsection{The faint branch as test particles}
\label{sec:test_particles}

\begin{figure*}[!htb]
\centering
\gridline{
\fig{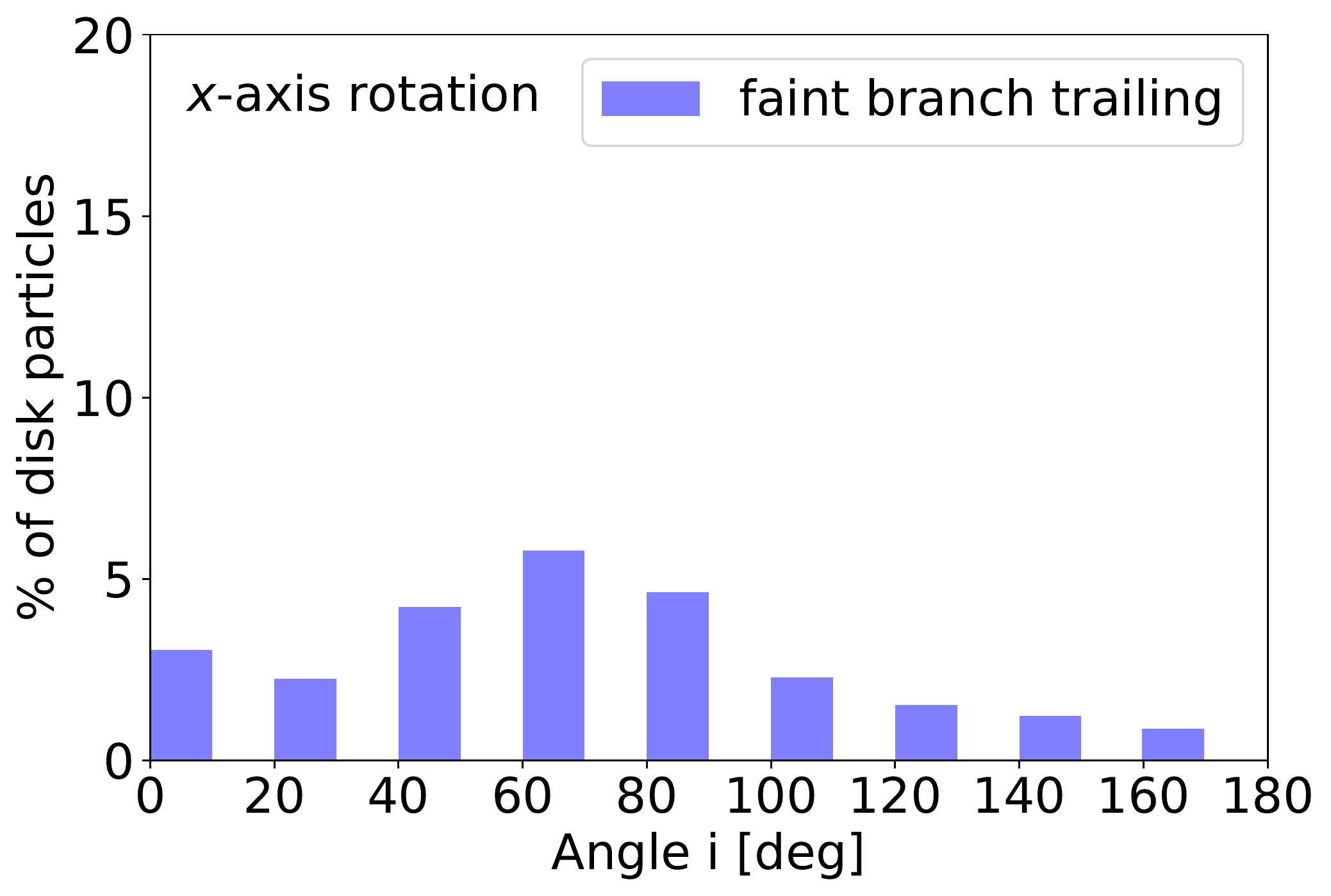}{0.32\textwidth}{}
\fig{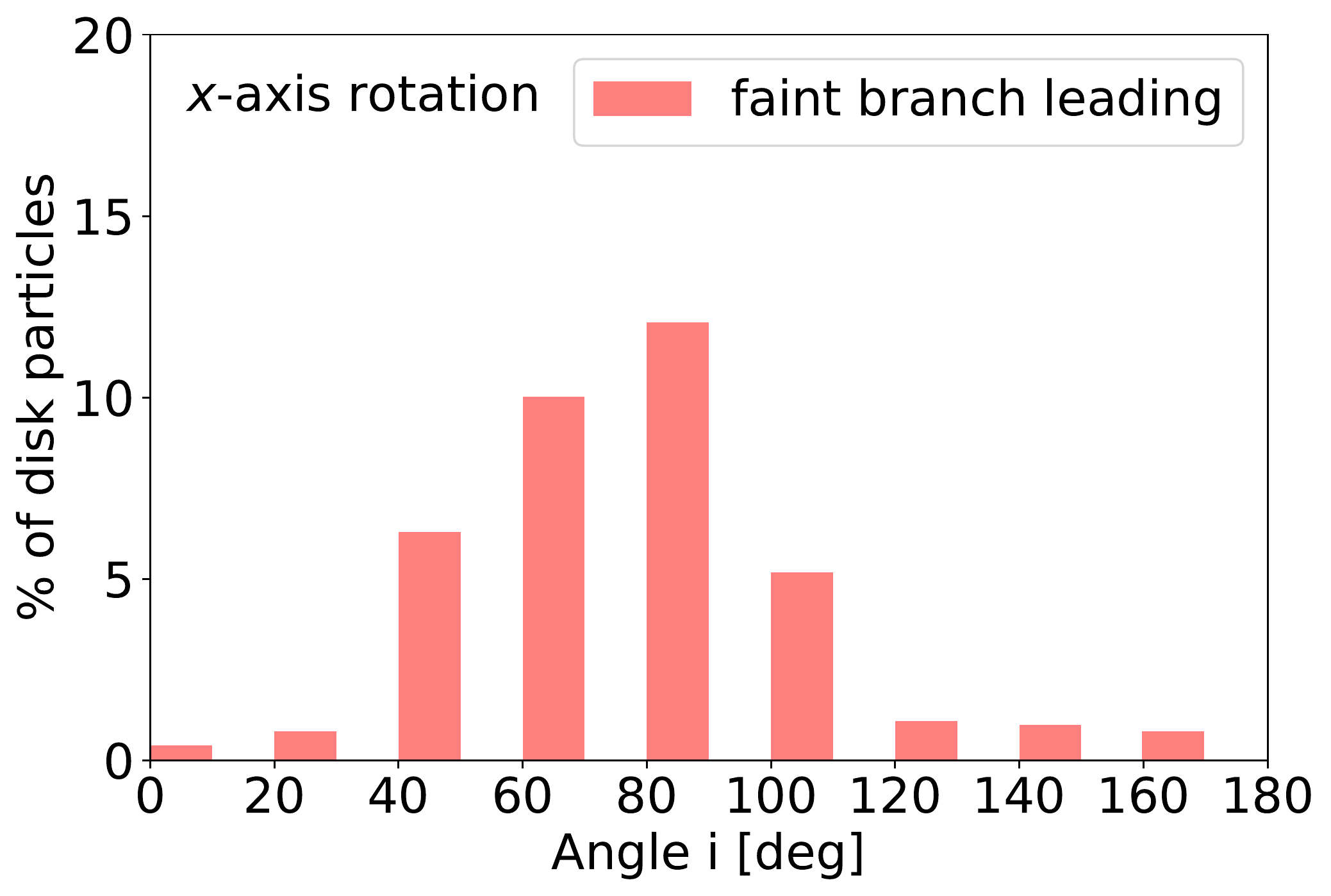}{0.32\textwidth}{}
\fig{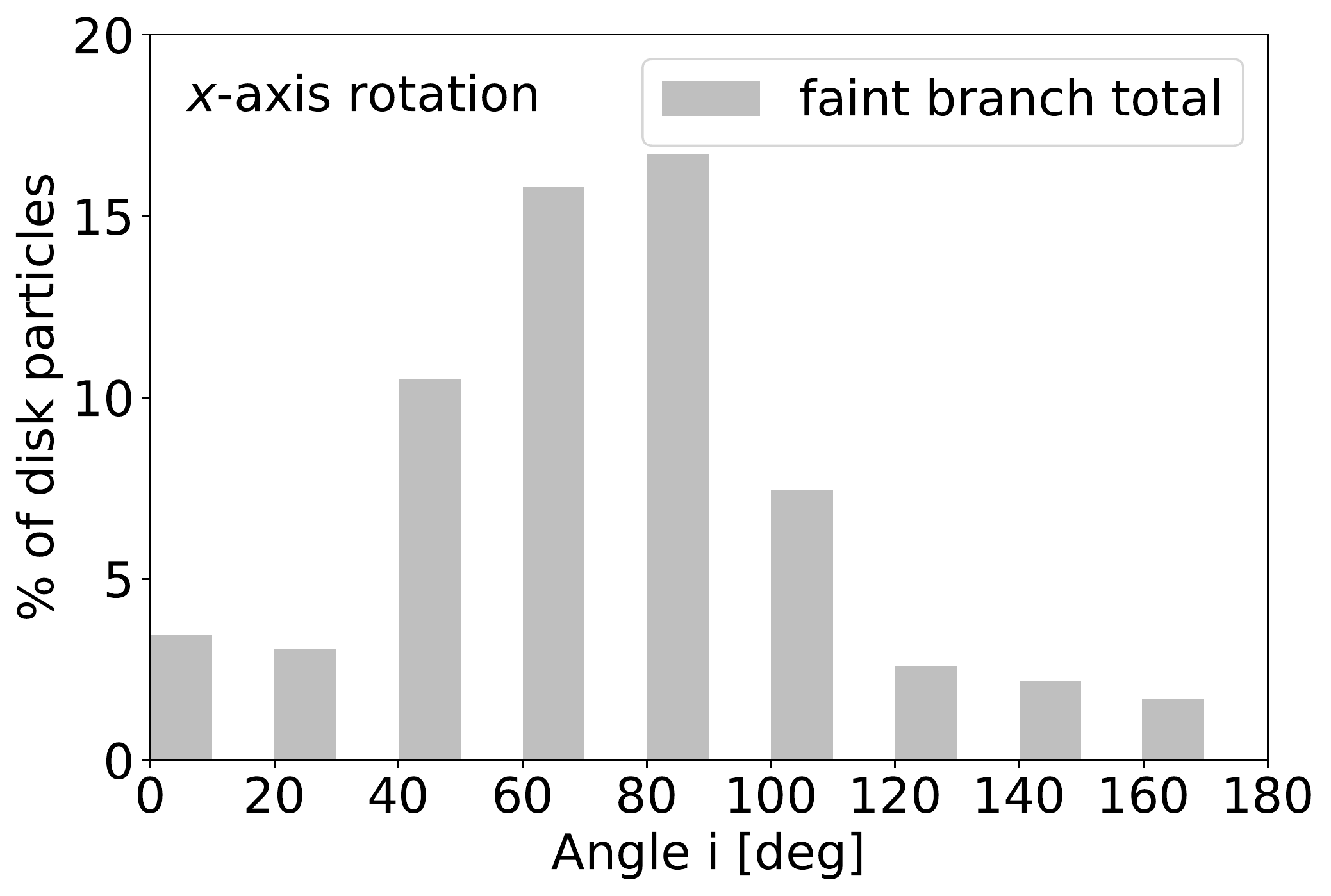}{0.32\textwidth}{}
}
\gridline{
\fig{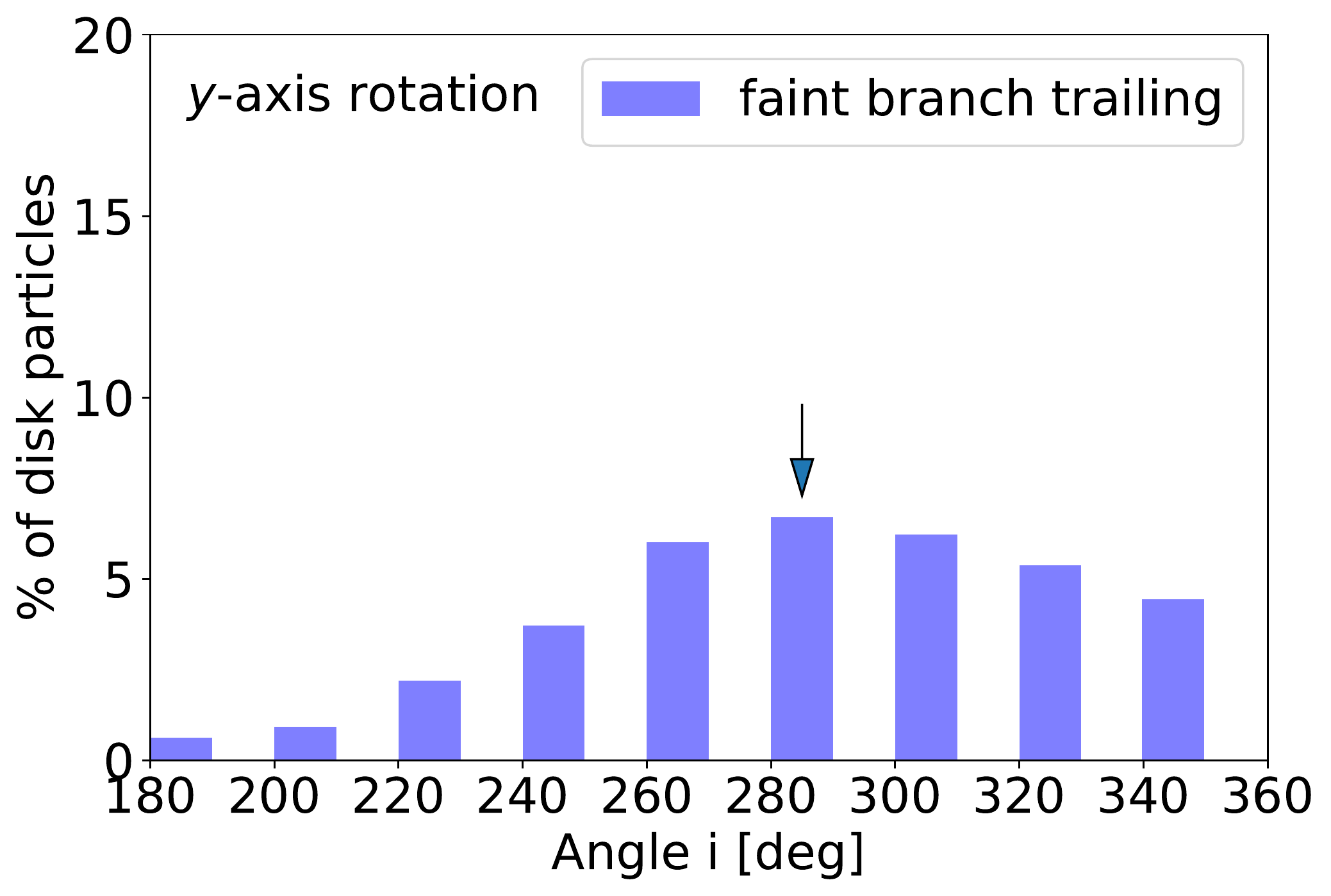}{0.32\textwidth}{}
\fig{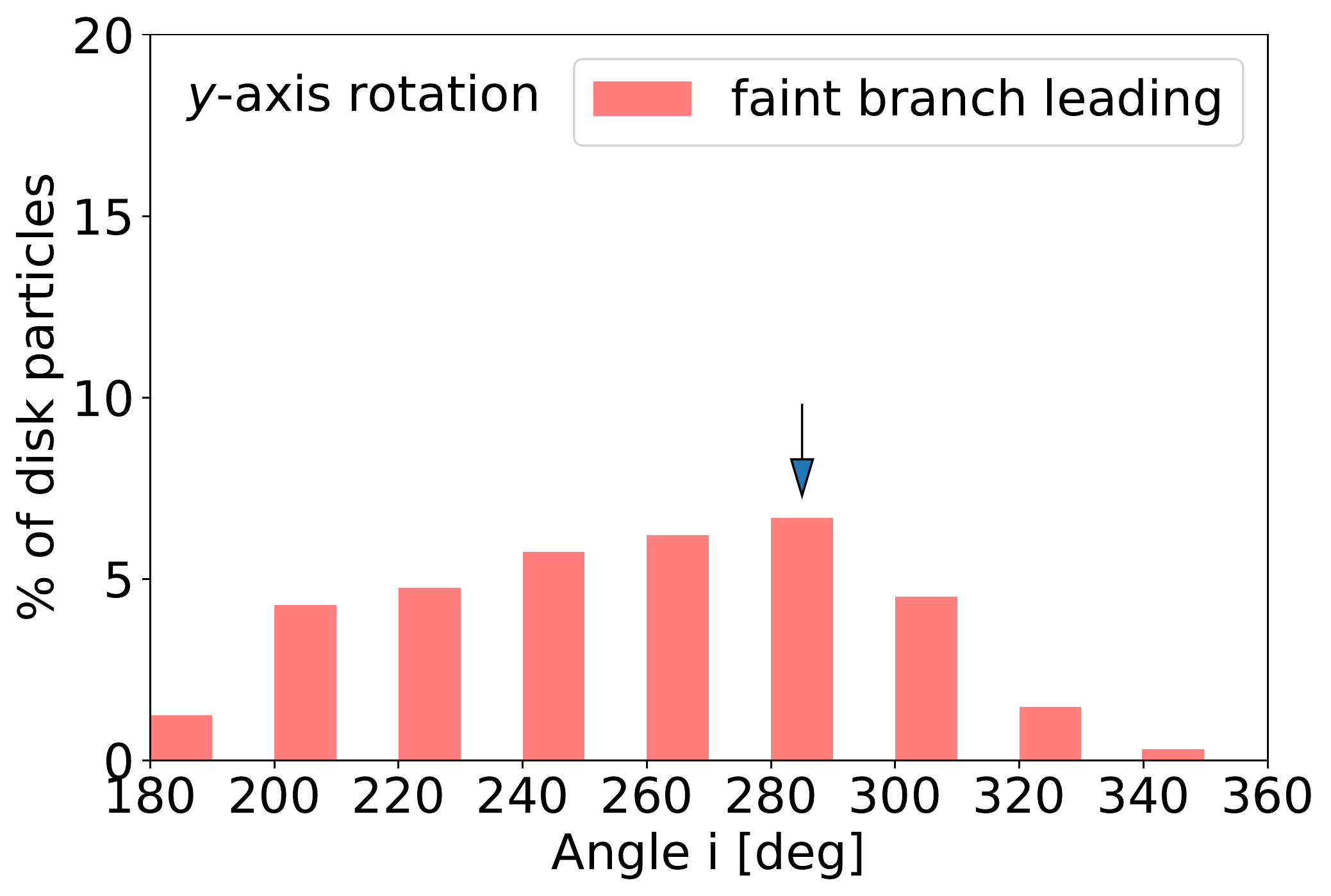}{0.32\textwidth}{}
\fig{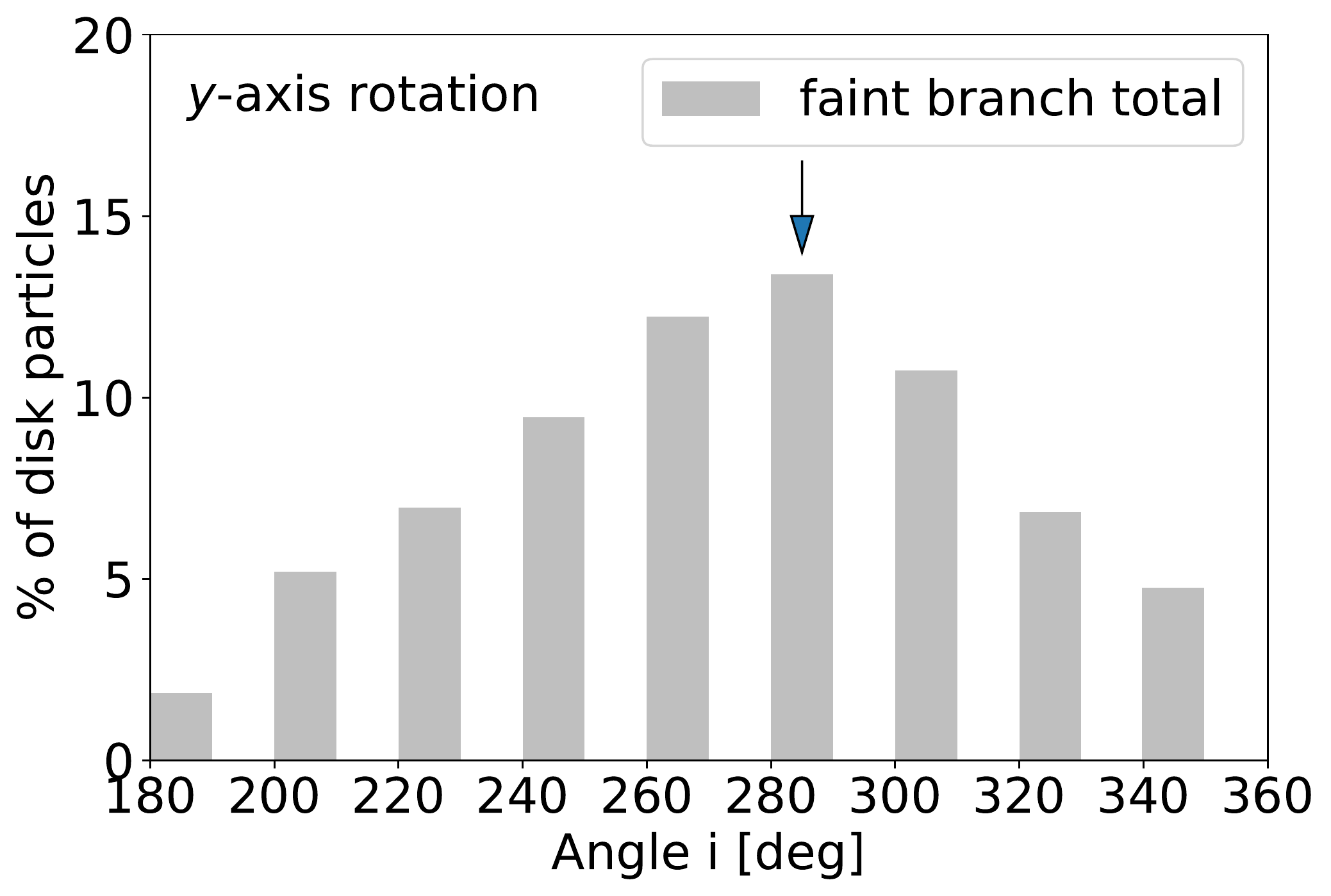}{0.32\textwidth}{}
}
   \caption{Fraction of disk particles that end up in the faint branch for a given disk inclination $i$. Each inclination angle is probed by a holed disk made of the $2\times10^4$ outermost test particles, with the holed disk rotated as explained in Section~\ref{sec:Methodology} around the $x$-axis (Sun-MW center axis) for the top panels, and around the $y$-axis for the bottom panels. The arrow points to our best model.}
\label{fig:angles}
\end{figure*}

We trace our selection of disk particles in the faint stream  (see Equations~\eqref{eq:polynomial_leading} and~\eqref{eq:polynomial_trailing} and Figure~\ref{fig:gaia_plus_test_particles}) back to the initial conditions.

Figure~\ref{fig:angles} shows the fraction of test particles that end up in the faint branch based on initial disk inclination and rotation. We only show the exploitable results: disks rotated around the $x$-axis by an angle $0^\circ\leq i \leq 180^\circ$ (top-panels) and disks rotated around the $y$-axis by an angle $180^\circ\leq i \leq 360^\circ$ (bottom-panels). We find that other rotations and angles do not lead efficiently to the creation of a faint branch, with at best $\simeq2-3$\% of particles ending in the desired regions.

\subsubsection{Best model: rotation around the $y$-axis}
\label{subsec:yaxis}

The most appealing model consists of a single disk rotated by an angle $i=280^\circ$ around the $y$-axis, which leads to a faint branch with roughly the same amount of stars in the leading and trailing parts, as can be seen in Figure~\ref{fig:angles}. We pick this option as the preferred model in this work. In this case, the disk almost lies in the $yz$ MW plane, making it nearly perpendicular both to the MW disk ($\sim xy$ MW plane) and to the Sgr orbital plane ($\sim xz$ MW plane).

\begin{figure*}[!htb]
\centering
  \subfigure{\includegraphics[angle=0,  clip, width=0.31\textwidth]{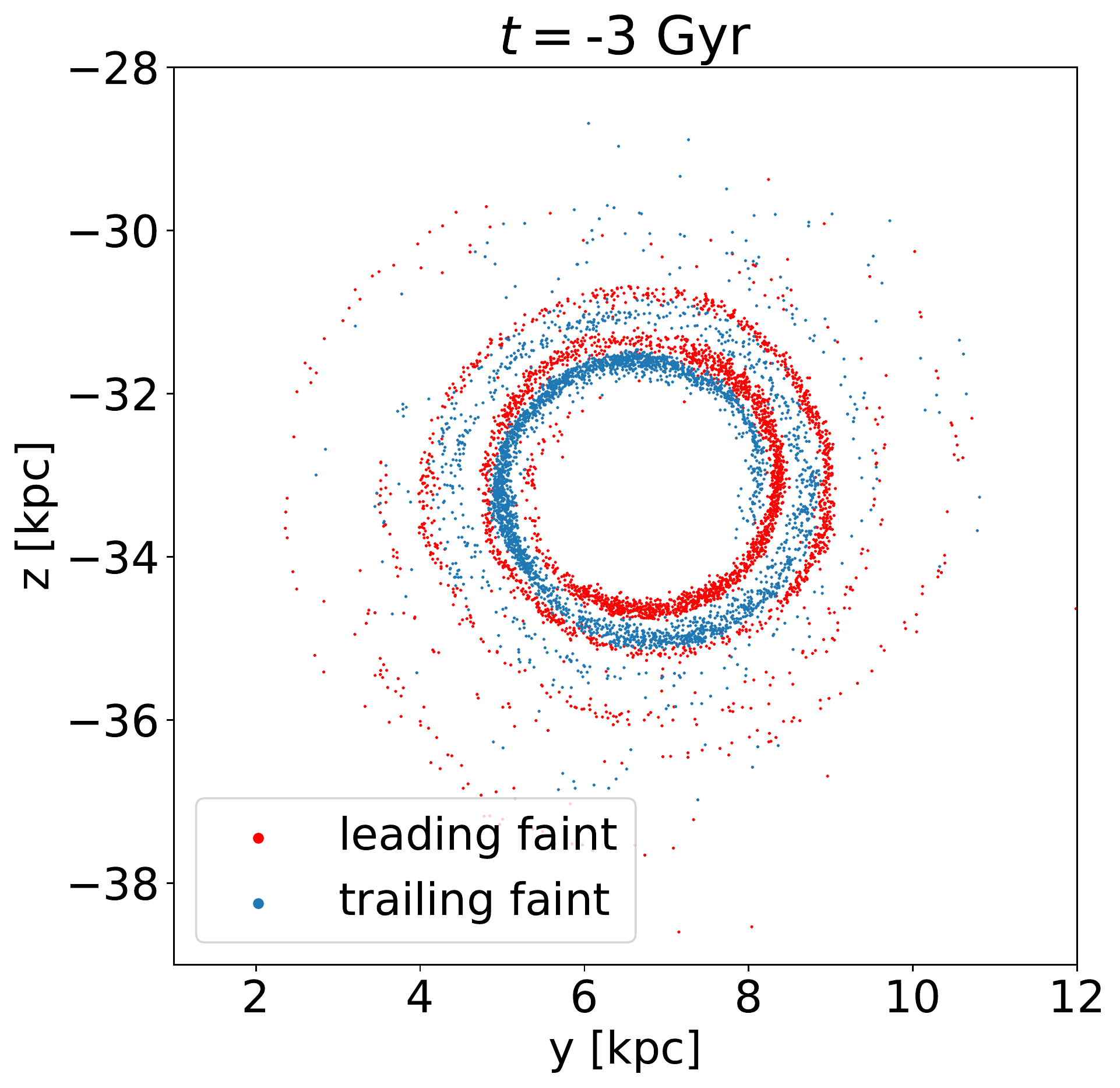}}
  \subfigure{\includegraphics[angle=0,  clip, width=0.32\textwidth]{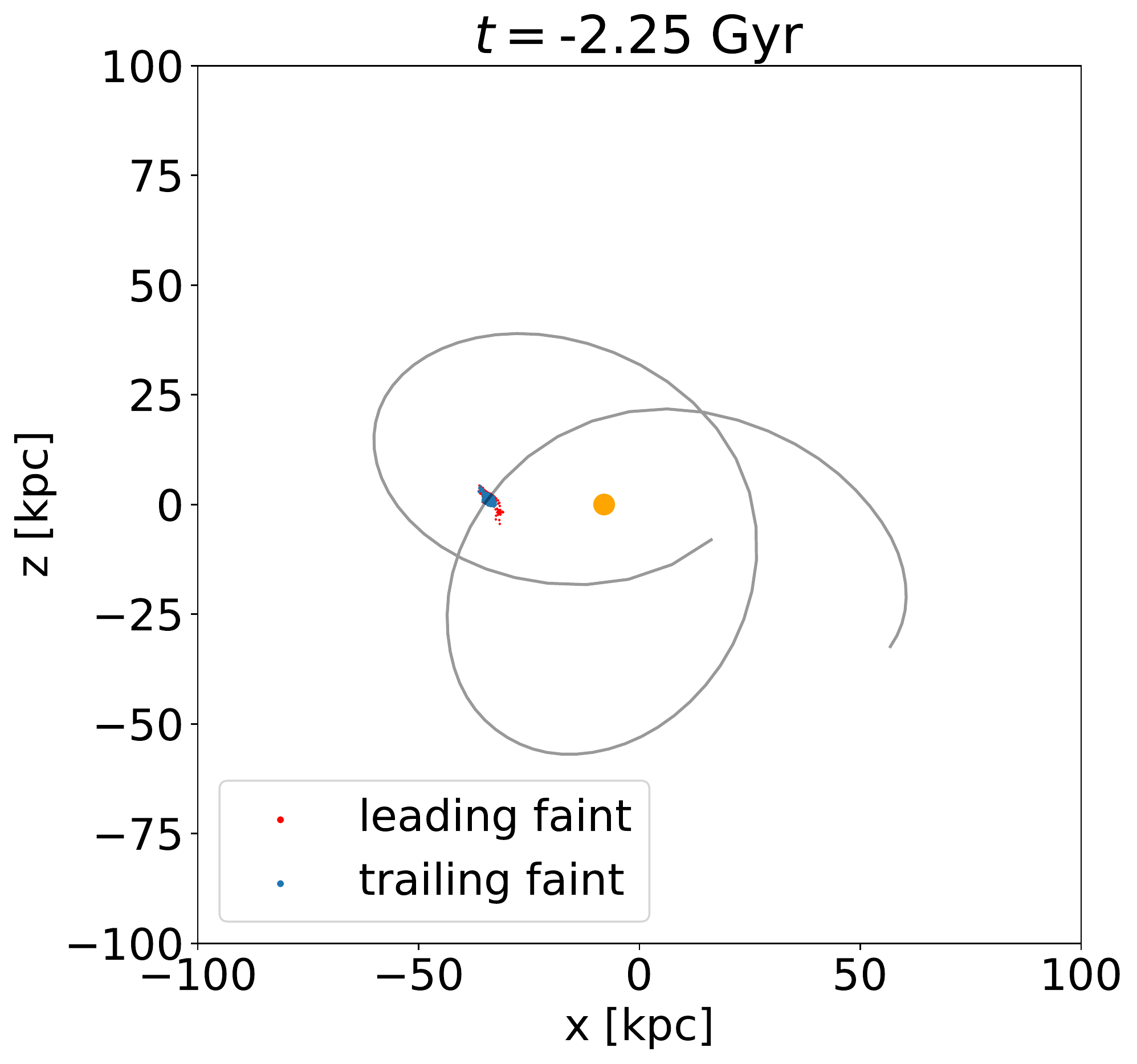}}
  \subfigure{\includegraphics[angle=0,  clip, width=0.32\textwidth]{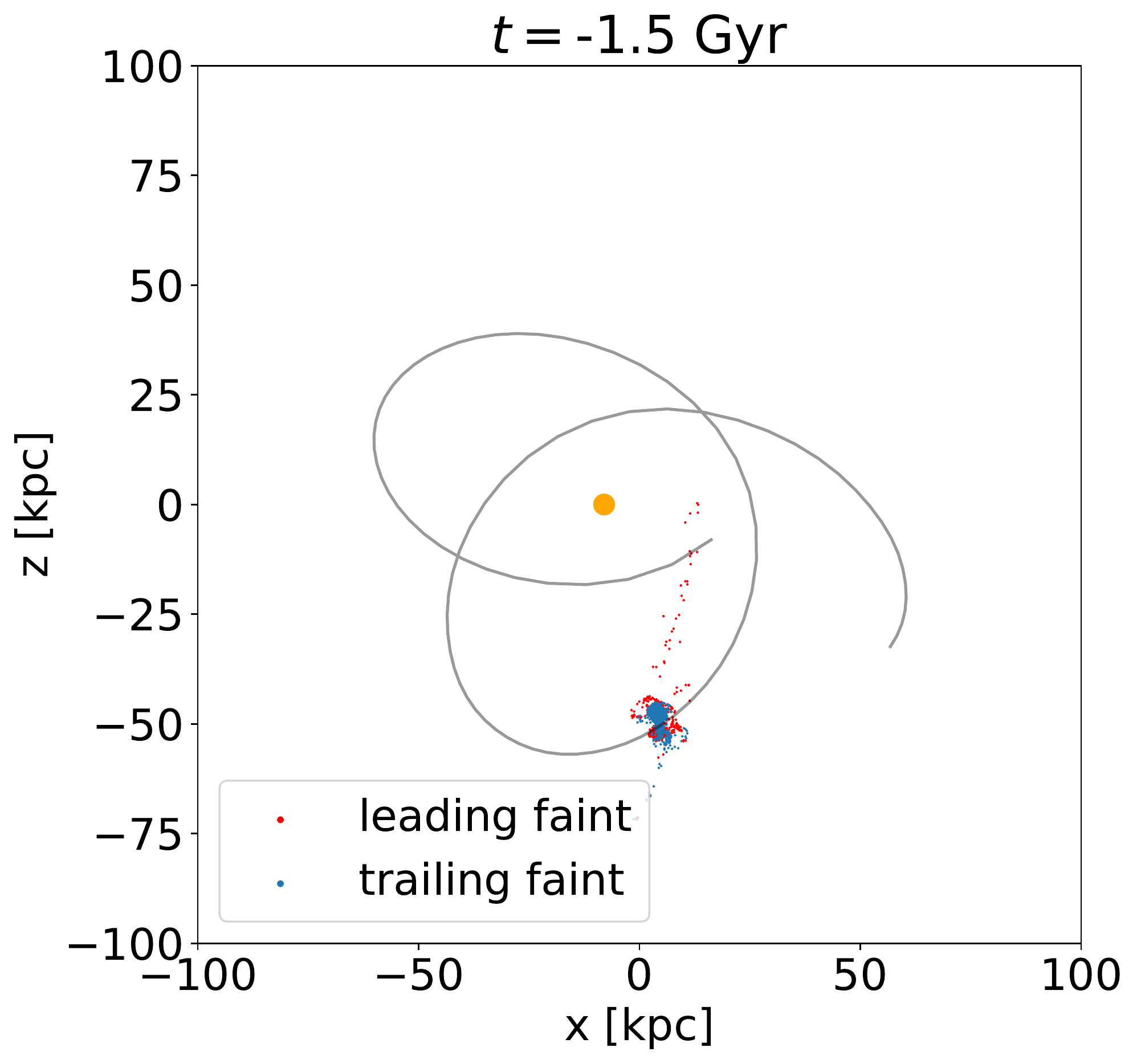}}
  \subfigure{\includegraphics[angle=0,  clip, width=0.32\textwidth]{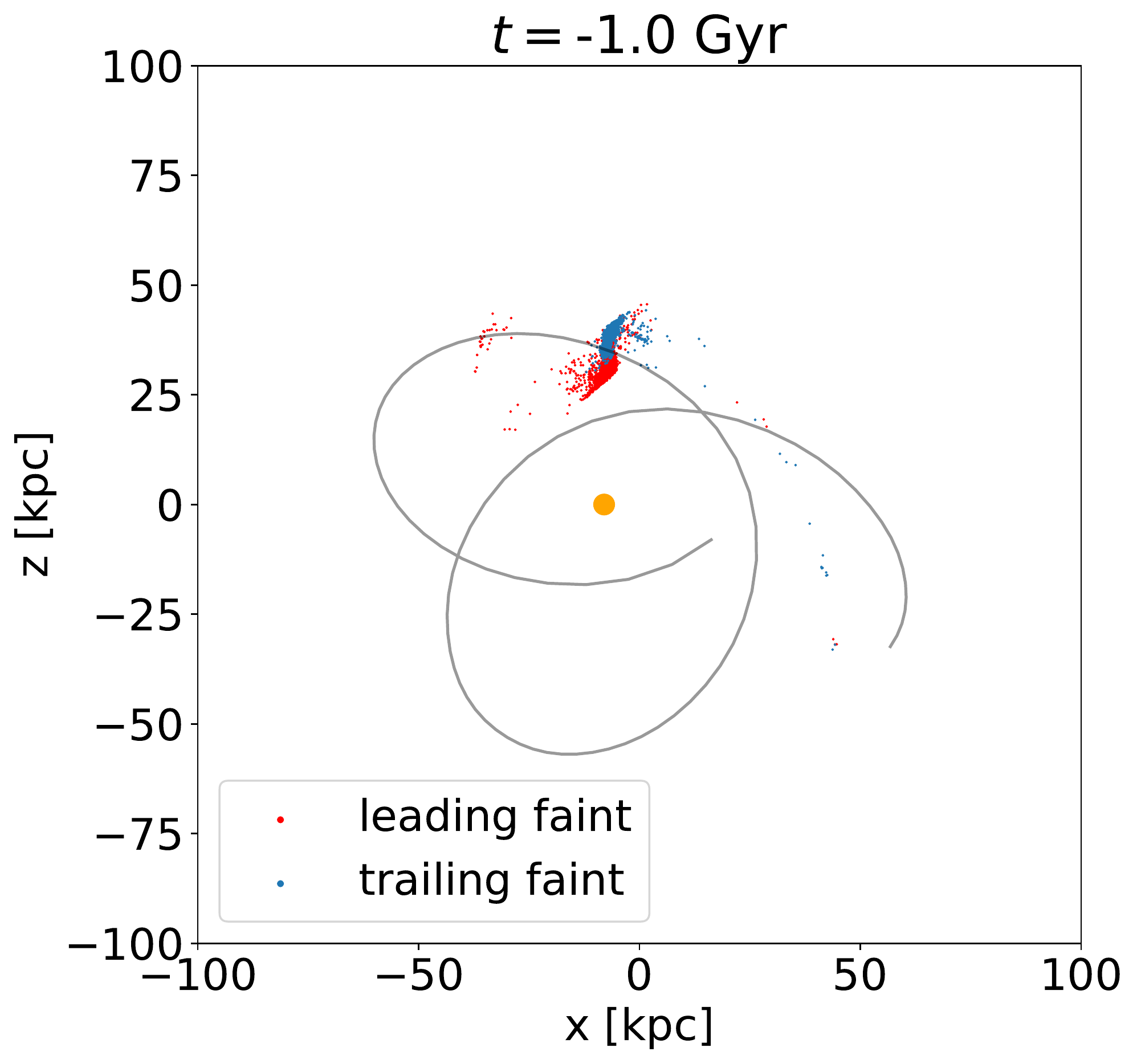}}
  \subfigure{\includegraphics[angle=0,  clip, width=0.32\textwidth]{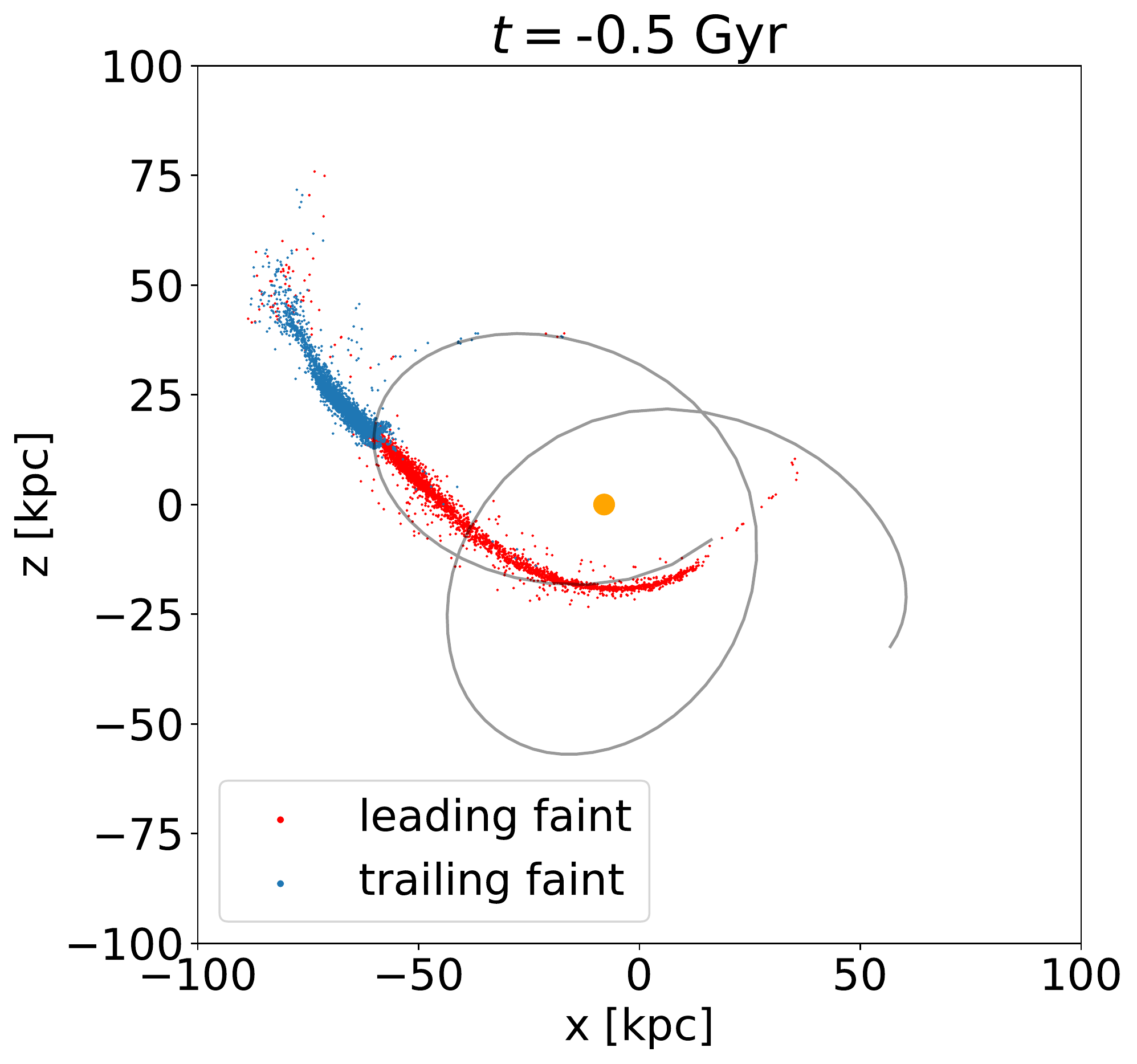}}
  \subfigure{\includegraphics[angle=0,  clip, width=0.32\textwidth]{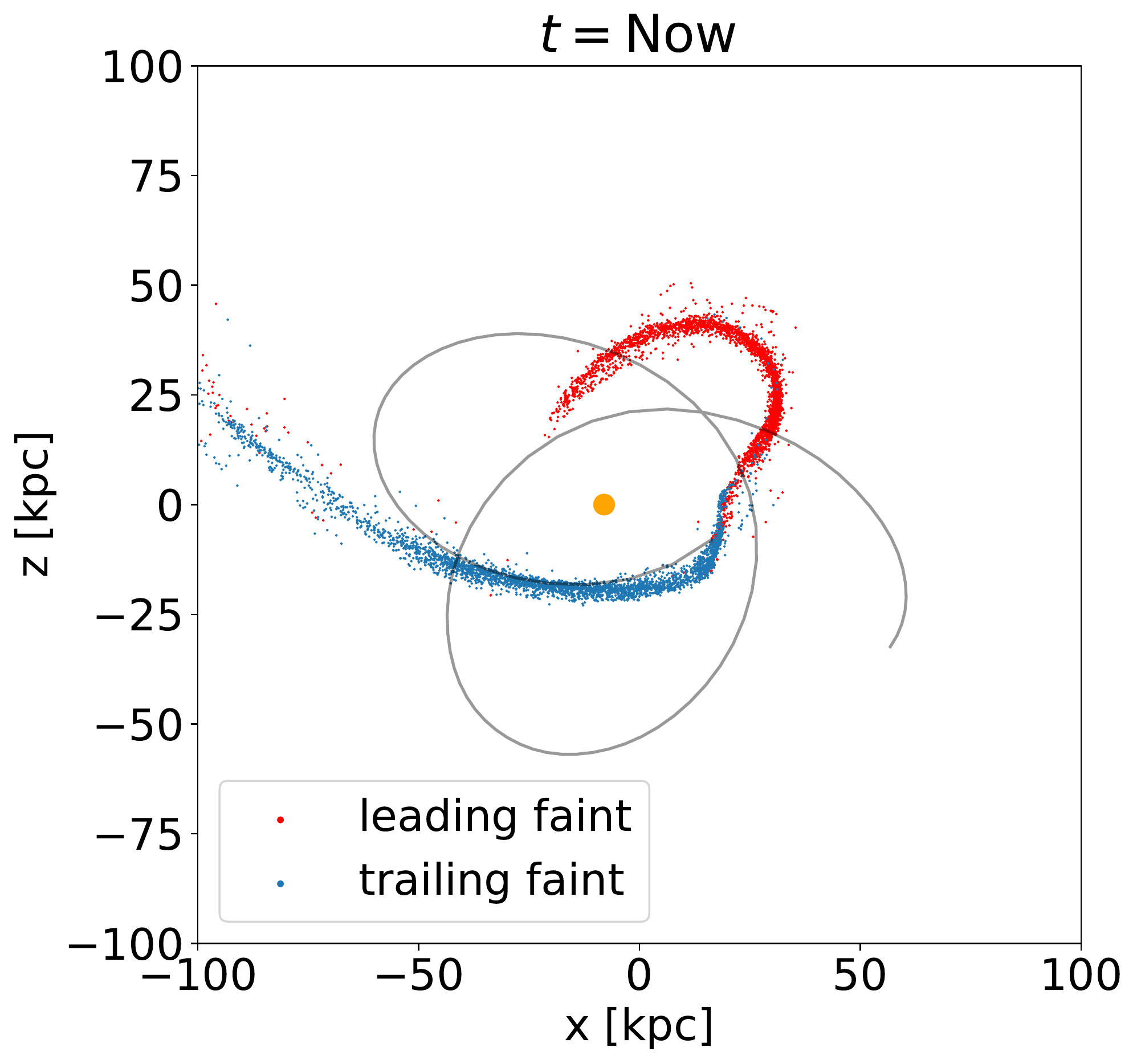}}
   \caption{Evolution of the faint branch selection, from spirals at the start of the simulation 3 Gyr ago (top-left panel, in the initial disk plane, close to the $yz$ MW plane) to present time (bottom-right panel). The evolution is seen in the $xz$ MW plane, close to the Sgr orbital plane. Particles that will make up the final leading arm are shown in red, and the final trailing arm in blue. The black curve represents the progenitor's orbit, and the orange ball represents the Sun. Pericentric passages of the Sgr occur around $t\simeq-2.3$ Gyr and $t\simeq-1.1$ Gyr. Similar plots for every snapshot of the simulation and the corresponding video are available in the shared data.}
\label{fig:spiral_evolution}
\end{figure*}

We thus run another simulation with the single disk rotated by an angle $i=280^\circ$ around the $y$-axis added to the model of \citetalias{V21}, still as test particles, but keeping this time the full disk, made of $10^5$ particles. After this rotation, our disk has angular momentum nearly aligned with the positive $x$-axis, with a small positive $z$ component. A majority of this initially full disk ends up in the bright branch or near the progenitor, and is not part of our selection. However, picking the disk test particles that end up in the faint branch once more, we are now interested in their distribution in the initial conditions. We find that our selection picks out high energy and angular momentum particles of the disk, and traces spiral arm-like features (Figure~\ref{fig:spiral_evolution}, top-left panel) in the outer disk, which would be sufficient to lead to the creation of the faint branch. 

In order to highlight the importance of angular momentum, we note in passing that doing the same exercise (selecting the faint branch and looking back in initial conditions) with the stellar particles of the King model of \citetalias{V21} does not lead to any clear signature distribution in position, energy, or angular momentum. 

\subsubsection{Alternative model: rotation around the $x$-axis}

Disks rotated around the $x$-axis with inclination angles $i=60^{\circ}$ and $i=80^{\circ}$ are also interesting, with $\simeq16-17$\% of the particles that end up in the faint branch (Figure~\ref{fig:angles}). This is not too surprising: at such inclinations, the disk plane matches closely the Sgr orbital plane. In this configuration, stars in the Sgr disk are on prograde orbits with respect to the orbit of Sgr around the MW. This has been shown in \citet{2015ApJ...810..100L} to lead to stars being stripped easily, producing thin streams. 

The model with a disk rotated by an angle $i=70^\circ$ around the $x$-axis is not implausible, but produces slightly worse results than our best disk model rotated by 280 degrees around the $y$-axis: the trailing arm is harder to populate, and the agreement with Gaia kinematics is not as good. From the Sgr orbital plane, the plane of such a disk makes an angle $-30^{\circ}$ around the $x$-axis. This value, which emerges naturally from our probing of the initial conditions when considering rotations around the $x$-axis, is very close to the value of $-20^{\circ}$ originally proposed by \citet{2010MNRAS.408L..26P}. Data and plots for this model are provided in the repository. 

\subsection{Including self-gravity}
\label{sec:self_gravity}

\begin{figure*}[!htb]
\centering
\includegraphics[angle=0,  clip, width=0.99\textwidth]{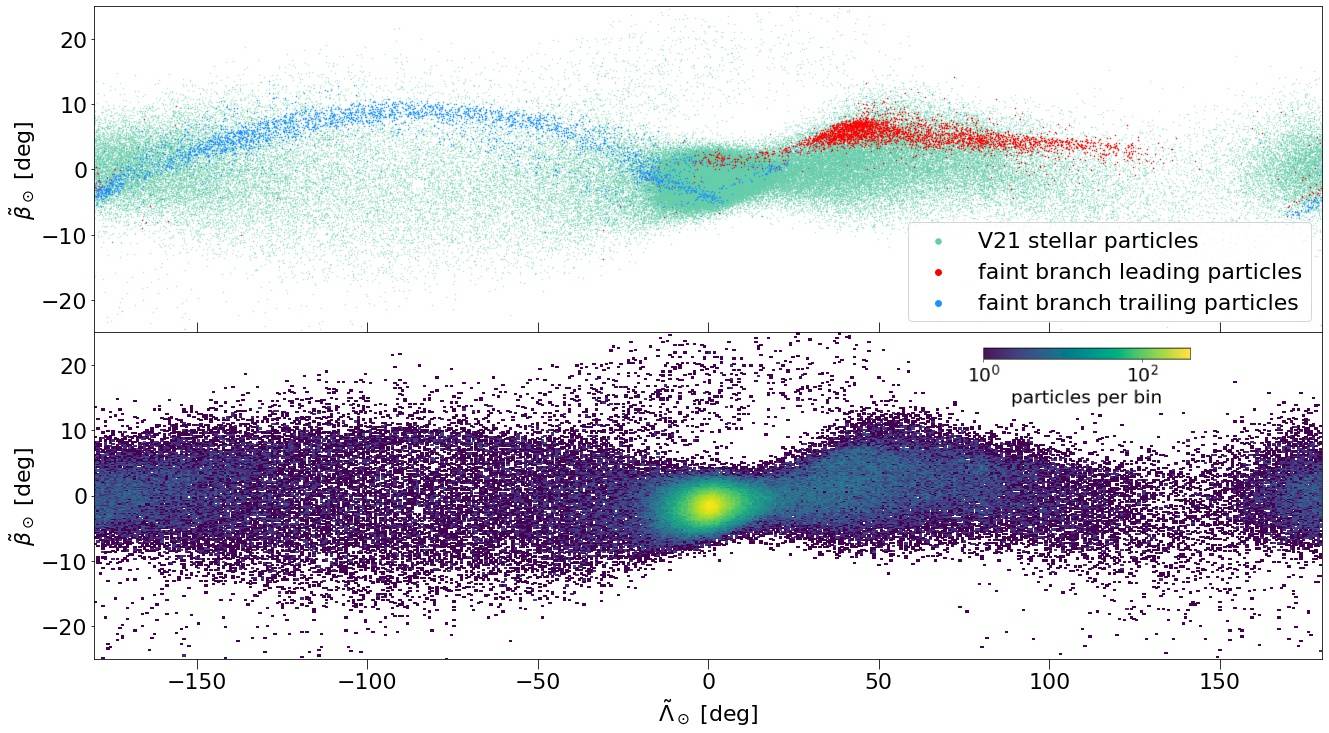}
     \caption{Final snapshot of the simulation (at present time) of our model with massive particles replacing some stellar particles of \citetalias{V21}'s initial model. \textit{Top panel:} The faint branch (red for leading arm, blue for trailing arm) is well populated over-plotted on \citetalias{V21}'s particles (green). \textit{Bottom panel:} Density plot of all stream stars (\citetalias{V21}'s stellar particles plus our faint branch particles).}
\label{fig:final_model_radec}
\end{figure*}

We now study whether the results of the previous section, obtained assuming that disk stars are mass-less tracers of the underlying potential, also hold when taking into account the self-gravity between disk stars. This will allow us to put forward a model that creates a faint branch like the one observed in the Sgr stream, using the initial conditions and gravitational potential of \citetalias{V21} as a backbone. 
In the Gaia EDR3 sample of Sgr of \citet{2021arXiv211202105R}, stars with probability $\geq80$\% of being part of the faint branch make up $\simeq4$\% of the total. We thus aim to be close to this ratio, and replace 6600 of the $2\times 10^5$ stellar particles in the \citetalias{V21} model by new ones following our initial disky distribution. In order to include our particles into the reference model, we give them the same mass as the stellar particles of \citetalias{V21}, and for each particle that we include, we remove one stellar particle from \citetalias{V21} sitting at the closest radius from that of our particle. Doing so ensures that we keep the same total mass, and does not alter the non-linear dynamics too much. 

We follow the evolution of our faint branch selection along the simulation in Figure~\ref{fig:spiral_evolution}, from initial spirals to eventually forming the faint branch. Our selection remains largely bound with angular momentum still pointing towards the positive $x$ direction until the second pericentric passage (around $t\simeq-1.1$ Gyr), which strips the faint branch particles from the progenitor.

Figure~\ref{fig:final_model_radec} (top panel) shows the $\tilde\Lambda_\odot$-$\tilde \beta_\odot$ view of the simulation at present time. The faint branch can be seen to be well populated, although with self-gravity now playing a role, a few of our particles end up close to the progenitor.

An issue is that the thick stream of the \citetalias{V21} model extends to the faint branch region already, resulting in an overly dense faint branch in Figure~\ref{fig:final_model_radec} (bottom panel). In a complete bifurcation model, the bright branch should ideally be thinner, which could probably be achieved with {\it e.g.} a different initial density profile, or a non-equilibrium transitional situation. We leave this exploration to a future study.

We compute the mean line-of-sight velocity in the remnant of the progenitor to make sure that our faint branch particles did not perturb the spherical model of \citetalias{V21} by adding significant rotation. We find a gradient of $\sim10$ km s$^{-1}$, similarly to the pressure-supported model of \citet[Fig. 2]{2011ApJ...727L...2P}. 

Interestingly, we note an over-density of our particles in the $-180^\circ\leq\tilde\Lambda_\odot\leq-130^\circ$  region ($70^\circ\leq\rm{RA}\leq120^\circ$), for which we do not have Gaia data. This signature appeared in all our simulations with disks of inclination close to that of the chosen model. It would be interesting to see if such an over-density can be observed.

\begin{figure*}[!htb]
\centering
\includegraphics[angle=0,  clip, width=\textwidth]{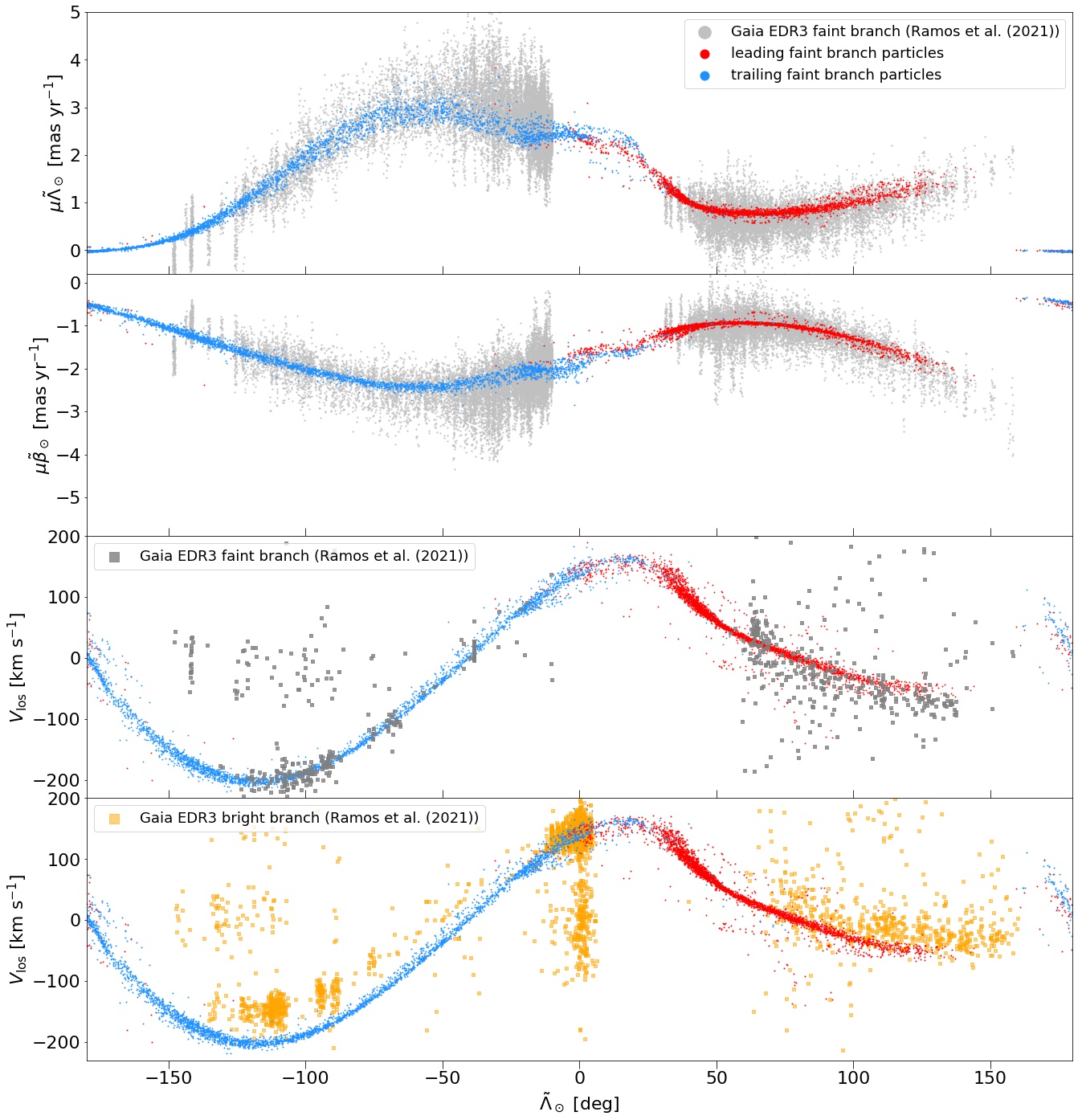}
  
   \caption{Comparison of the faint branch (probability $\geq80\%$) from the Gaia EDR3 Sgr sample of \citet[][grey]{2021arXiv211202105R} with the faint branch selection in our $N$-body model including self-gravity (red for leading arm, blue for trailing arm). From top to bottom: proper motions in $\tilde\Lambda_\odot$, in $\tilde \beta_\odot$, and line-of-sight velocities (heliocentric reference frame). For comparison, the bottom panel shows bright branch members (probability $\geq80\%$, yellow).}
\label{fig:final_model_astrometry}
\end{figure*}

Finally, we compare in Figure~\ref{fig:final_model_astrometry} our faint branch particles to the faint branch selection (probability $\geq80\%$) from the Gaia EDR3 sample of \citet{2021arXiv211202105R}. The radial velocities follow the observed trend for the faint branch in the trailing and leading arms, and are different from those of the bright branch in agreement with the data. We remind that our initial selection has been made purely in configuration space, so that this phase-space agreement is impressive. For proper motions, the difference in trends between the bright and faint branches is small in the data, as can be seen in \citet[Fig. 3]{2021arXiv211202105R}. We note in passing that our faint branch simulation has smaller scatter than the observed data in proper motions, probably due to both intrinsic dispersion (both in velocity and distance) and observational uncertainties. However, transforming our model into star particles with magnitudes and hence Gaia astrometric uncertainties is far beyond the scope of this contribution. 

\section{Discussion}

Despite the obvious similarities with a full disk model \citep{2010MNRAS.408L..26P}, there are a few differences. For one, this spiral disky distribution allows to populate only the faint branch and can be combined with a more massive spherical model to also populate the bright branch. It also alleviates the issue of requiring substantial rotation in the progenitor's remnant at present time \citep{2011ApJ...727L...2P}. While out of the scope of the present work to produce a full Sgr model, we discuss possibilities that would lead to the presence of our faint branch selection and how the inner Sgr could be populated.

A possible scenario for the presence of this disky spiral distribution 3~Gyr ago could be a disky dwarf perturbed by tidal effects \citep{2017ApJ...842...56G}, and/or having been affected by disk-shocking while crossing the MW-disk. In addition to the spirals, the rest of the disk could be transitioning to a pressure supported spheroidal galaxy as in the tidal stirring mechanism \citep{2001ApJ...547L.123M}. In this context, the inner galactic disk often goes through a bar perturbation \citep{2011ApJ...726...98K,2014MNRAS.445.1339L}. It is thus possible that a bar would be present in Sgr 3 Gyr ago. With tidal heating, the bar transforms into a diffuse spheroid, part of which would then end up in the bright branch of the stream, and the rest would form the elongated remnant of the progenitor that is now observed. This model is attractive because both branches of the Sgr stream would come from the same stellar population, consistent with the small difference observed in metallicity between the faint and bright branches \citep{2021arXiv211202105R}. 

Another possibility would be that Sgr was already having a substantial spheroidal component and that only a remaining rotating disk was affected. Indeed, fitting a full exponential density profile from the surface density profile of our spiral selection and extrapolating it to the inner disk, we find that a total disk mass of $2\sim3\times10^7$ M$_\odot$, or $10\sim15$\% of the mass of the stellar component in \citetalias{V21}'s model would be sufficient (the mass range depending on the proportion of faint branch stars, $\simeq4$\% of stream stars in the data). Such a minor disk component would produce a very low rotation signal in the Sgr remnant at present time.  

The spiral features could also be the tidal tails or stellar stream of an accreted globular cluster or dwarf galaxy inside the Sgr system. This is however less enticing as it would require the stellar populations of Sgr and the putative satellite to be fairly similar. 

It is also interesting to compare the stripping history and geometry of this faint branch to full models (see \textit{e.g.} \citealt[their Fig. 7]{2021arXiv211202105R}). As shown in Figure~\ref{fig:spiral_evolution} and in the shared material, our particles for the leading and trailing faint branch are both mostly stripped during the second pericenter of the simulation ($t\simeq-1.1$ Gyr). In addition, this stripping produces a single ``upper'' faint branch that can be paired with another Sgr component that would produce the parallel bright branch, as opposed to the undesired ``X-shape'' \citep{2021arXiv211202105R} that is usually obtained when considering inner rotation and/or orbital precession.

\section{Conclusion}
\label{sec:Conclusion}

We propose a model for the bifurcation of the Sgr stream in which the faint branch is populated by stars that were distributed in a disky spiral distribution within the progenitor 3~Gyr ago, in a plane nearly perpendicular to both the Sgr orbital plane and the MW disk plane. This pattern emerged here naturally by probing a large range of initial position, energy and angular momentum distributions for stellar test particles that end up in both the leading and trailing parts of the observed faint branch. Populating the faint branch this way opens the possibility of freely pairing this work with other Sgr components that would produce the parallel bright branch.

In the context of the tidal stirring mechanism studied in detail in \citet{2011ApJ...726...98K} for the formation of dwarf spheroidals, Sgr could previously have been a disky galaxy which 3 Gyr ago already held a bar \citep{2014MNRAS.445.1339L} and started the transition from a disky galaxy to a more isotropic and diffuse one. Low rotational velocity would then be observed today in the remnant, in agreement with \citet{2021ApJ...908..244D}. The spiral features could be tidally-induced, bar-induced, or the result of disk shocking when crossing the MW. Although out of scope for the present study, it would be very interesting to see if this could be turned into a working model for the entire Sgr stream. 

Another interesting albeit less likely possibility would be that this spiral distribution is the tidal tail or stellar stream caused by the disruption of a satellite of the Sgr system. Further observations of the stellar populations and their detailed chemistry in both the bright and faint branch will likely provide very useful information in deciding this matter. 

\section*{Data Availability}
\label{sec:Data}
We make available our addition to the model of \citetalias{V21} (\href{https://doi.org/10.5281/zenodo.4300977}{DOI 10.5281/zenodo.4300977}), including the particles leading to the creation of the faint branch as well as plots, a movie, and all related scripts, at \href{https://doi.org/10.5281/zenodo.6581185}{DOI 10.5281/zenodo.6581185}.

\begin{acknowledgments}
The authors thank the referee for an extremely helpful report which led to significant improvements of the paper. The authors acknowledge funding from the European Research Council (ERC) under the European Unions Horizon 2020 research and innovation programme (grant agreement No. 834148) and from the Agence Nationale de la Recherche (ANR projects ANR-18-CE31-0006 and ANR-19-CE31-0017).

\end{acknowledgments}

\bibliography{sgr_bifurc}
\bibliographystyle{aasjournal}

\end{document}